\documentclass[12pt]{article}
\usepackage{epsf}
\usepackage[utf8]{inputenc}
\usepackage[english]{babel}

\begin{document}

\title{
{\bf High-energy single diffractive dissociation of nucleons and the 3P-model applicability range}}
\author{A.A. Godizov\thanks{E-mail: anton.godizov@gmail.com}\\
{\small {\it SRC Institute for High Energy Physics of NRC ``Kurchatov Institute'', 142281 Protvino, Russia}}}
\date{}
\maketitle

\vskip-1.0cm

\begin{abstract}
The adequacy of the triple-Pomeron interaction approximation (the 3P-model) for description of the high-energy single diffractive dissociation of nucleons is analyzed via 
application to the available experimental data on nucleon-nucleon scattering, including the recent results produced by CMS Collaboration which allow to estimate reliably 
the triple-Pomeron coupling value. It is argued that the total contribution of secondary Reggeon exchanges is not negligible up to the Tevatron energy.
\end{abstract}

\section*{Introduction}

The reaction of high-energy single dissociation of proton, $p+p\to p+X$ or $\bar p+p\to \bar p+X$, where $X$ is the produced hadronic state of mass $M_X\ge m_p+m_{\pi^0}$, is 
usually called diffractive if the energy fraction lost by the surviving particle, $\xi=(M_X^2-m_p^2)/s$, does not exceed 0.05. More than 10\% of all the proton-(anti)proton 
collision events at the SPS, Tevatron, and the LHC are single-diffractive. Hence, the importance of the single diffraction (SD) studies at high-energy colliders is evident. 

The most natural theoretical framework for description of the SD observables is Regge theory \cite{collins}, and, in particular, the triple-Pomeron interaction approximation 
(the 3P-model) for high values of the missing mass ($M_X^2\gg m_p^2$). The Regge formalism is widely used in the papers on SD phenomenology \cite{fiore}---\cite{shuvaev}. 
However, regarding the applications of the Pomeron exchange approximation to the high-energy SD processes, such a problem emerges as the presence of secondary Reggeon 
exchanges ({\it i.e.}, the necessity of unambiguous determination of the 3P-model applicability range), and this problem is still far from its final solution. The aim of this 
eprint is to reanalyze the 3P-model practical relevance in view of the recent experimental data produced by the CMS Collaboration \cite{CMS}, as well as to obtain (with the 
help of these data) a more reliable estimation of the triple-Pomeron coupling value.

\section*{The 3P-interaction approximation {\it versus} available experimental data}

Usually, high-energy elastic and single diffraction reactions are treated in terms of the invariant Mandelstam variables $s=(p_1+p_2)^2$ and $t=(p_1-p_1')^2$, where $p_1$ and 
$p_2$ are the 4-momenta of the incoming particles and $p_1'$ is the 4-momentum of the scattered (anti)proton. However, if we take account of the in-channel absorption 
(the significance of absorptive corrections in SD is discussed in detail in \cite{maor} and \cite{martynov}), then it is more convenient to use such a quantity as 
the transverse transferred momentum $\vec\Delta_\perp$ associated with $t$ by means of the relation\linebreak 
$t=-(\vec\Delta_\perp^2+\xi^2m_p^2)/(1-\xi)+O(1/s)\approx-\vec\Delta_\perp^2$ at $\xi\ll 1$. Below we omit the subscript ``$\perp$'' and consider all the nonscalar variables 
as vectors orthogonal to the beam axe.

A reliable way to estimate the impact of secondary Reggeon exchanges on the SD cross-sections is to fit the 3P-model degrees of freedom to the experimental data 
in that kinematic range wherein the total contribution of secondaries is negligible and, then, to compare the model predictions with other available data.

\begin{figure}[ht]
\begin{center}
\epsfxsize=12cm\epsfysize=9cm\epsffile{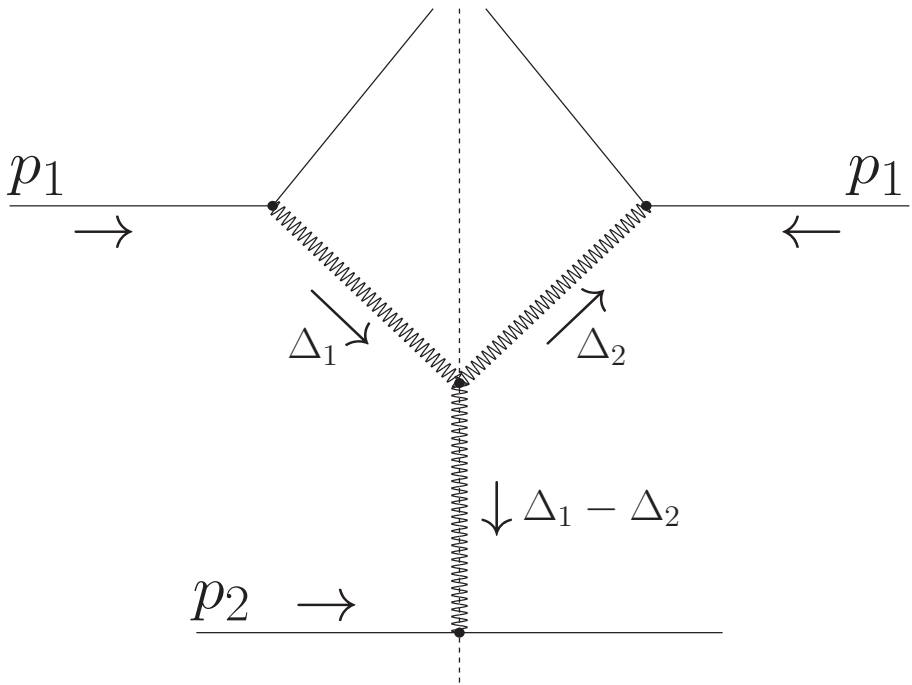}
\end{center}
\vskip-0.7cm
\caption{The triple-Pomeron interaction diagram.}
\label{triple}
\end{figure}
First of all, let us formulate concisely the 3P-model itself (the detailed consideration can be found in \cite{collins}). The ground of the 3P-approximation is the Mueller 
generalized optical theorem \cite{mueller} which allows to relate the SD observables in the asymptotic kinematic regime\linebreak $\sqrt{s}\gg M_X\gg \{m_p,|\vec\Delta|\}$ to 
the triple-Pomeron interaction amplitude (see Fig. \ref{triple}):
$$
T_{\rm 3P}(s\,,\,M_X\,,\,\vec\Delta_1\,,\,\vec\Delta_2) = \frac{1}{s}\left(i+\tan\frac{\pi(\alpha_{\rm P}(-\vec\Delta_1^2)-1)}{2}\right)
\left(-i+\tan\frac{\pi(\alpha_{\rm P}(-\vec\Delta_2^2)-1)}{2}\right)\times
$$
$$
\times\;g_{pp\rm P}(-\vec\Delta_1^2)\;g_{pp\rm P}(-\vec\Delta_2^2)\;g_{pp\rm P}(-|\vec\Delta_1-\vec\Delta_2|^2)\;
g_{3\rm P}(-\vec\Delta_1^2\,,\,-\vec\Delta_2^2\,,\,-|\vec\Delta_1-\vec\Delta_2|^2)\;\times
$$
\begin{equation}
\label{tripamp}
\times\;\pi^3\alpha'_{\rm P}(-\vec\Delta_1^2)\,\alpha'_{\rm P}(-\vec\Delta_2^2)\,\alpha'_{\rm P}(-|\vec\Delta_1-\vec\Delta_2|^2)\;\times
\end{equation}
$$
\times\;\left(\frac{s}{2s_0}\right)^{\alpha_{\rm P}(-\vec\Delta_1^2)+\alpha_{\rm P}(-\vec\Delta_2^2)}
\left(\frac{M_X^2}{2s_0}\right)^{\alpha_{\rm P}(-|\vec\Delta_1-\vec\Delta_2|^2)-\alpha_{\rm P}(-\vec\Delta_1^2)-\alpha_{\rm P}(-\vec\Delta_2^2)}\sim
$$
$$
\sim\frac{1}{s}\left(\frac{1}{\xi}\right)^{\alpha_{\rm P}(-\vec\Delta_1^2)+\alpha_{\rm P}(-\vec\Delta_2^2)}
M_X^{2\alpha_{\rm P}(-|\vec\Delta_1-\vec\Delta_2|^2)}\,,
$$
where $s_0=1$ GeV$^2$, $\alpha_{\rm P}(t)$ is the Regge trajectory of the Pomeron, $g_{pp\rm P}(t)$ is the Pomeron coupling to nucleon, and the symmetric function 
$g_{3\rm P}(t_1,t_2,t_3)$ is the 3P-interaction vertex. Such factors as $\pi\alpha'_{\rm P}$ and $2^{-\alpha_{\rm P}}$ are singled out within the Regge residue for 
the same reasons as for the elastic scattering \cite{godizov}.

If we introduce the Fourier-image of this function, 
\begin{equation}
\label{fourier}
T_{\rm 3P}(s\,,\,M_X\,,\,\vec b_1\,,\,\vec b_2)\equiv\frac{1}{(16\pi^2s)^2}
\int d^2\vec\Delta_1 d^2\vec\Delta_2\;e^{i\vec\Delta_1\vec b_1}\;e^{-i\vec\Delta_2\vec b_2}\;T_{\rm 3P}(s\,,\,M_X\,,\,\vec\Delta_1\,,\,\vec\Delta_2)\,,
\end{equation}
then the invariant one-particle inclusive cross-section (with account of absorption) can be represented as \cite{eden}
\begin{equation}
\label{diffcross}
16\pi^2s\frac{d^2\sigma}{dt dM_X^2} = (4s)^2\int d^2\vec b_1d^2\vec b_2\;e^{-i\vec\Delta\vec b_1}e^{i\vec\Delta\vec b_2}
\left[e^{i\delta(s,b_1)}\,T_{\rm 3P}(s\,,\,M_X\,,\,\vec b_1\,,\,\vec b_2)\,e^{-i\delta^*(s,b_2)}\right]\,,
\end{equation}
where $\delta(s,b)$ is the high-energy elastic scattering eikonal (Born amplitude) \cite{godizov} in the impact parameter representation:
\begin{equation}
\label{eik}
\delta(s,b) = \frac{1}{16\pi^2 s}\int d^2\vec\Delta\;e^{i\vec\Delta\vec b}\;\delta(s,-\vec\Delta^2) = \frac{1}{16\pi s}\int_0^{\infty}d(\vec\Delta^2)\,
J_0(b\,|\vec\Delta|)\,\delta(s,-\vec\Delta^2) = 
\end{equation}
$$
= \frac{1}{16\pi s}\int_0^{\infty}d(-t)\,J_0(b\sqrt{-t})\,\left(i+\tan\frac{\pi(\alpha_{\rm P}(t)-1)}{2}\right)
g^2_{pp\rm P}(t)\;\pi\alpha'_{\rm P}(t)\left(\frac{s}{2s_0}\right)^{\alpha_{\rm P}(t)}\,.
$$

For making quantitative predictions we need to fix the model degrees of freedom, namely, the unknown functions $\alpha_{\rm P}(t)$, $g_{pp\rm P}(t)$, and 
$g_{3\rm P}(t_1,t_2,t_3)$. The Pomeron Regge trajectory and the Pomeron coupling to nucleon should be the same as in the elastic scattering \cite{godizov}, 
\begin{equation}
\label{pomeron}
\alpha_{\rm P}(t) = 1+\frac{\alpha_{\rm P}(0)-1}{1-\frac{t}{\tau_a}}\,,\;\;\;\;g_{pp\rm P}(t)=\frac{g_{pp\rm P}(0)}{(1-a_gt)^2}\,,
\end{equation}
where the free parameters take on the values presented in Table \ref{tab1}.

Such a choice of parametrization for $\alpha_{\rm P}(t)$ is specified by the following from QCD asymptotic behavior of the Pomeron Regge trajectory \cite{low,kearney}, 
\begin{equation}
\label{asyP}
\lim_{t\to-\infty}\alpha_{\rm P}(t)=1\,,
\end{equation}
and by the conditions 
\begin{equation}
\label{deriv}
\frac{d^n\alpha(t)}{dt^n}>0\;\;\;\;(n=1,2...\,;\;t<0)
\end{equation}
which originate from the dispersion relations for Regge trajectories \cite{collins} and are expected to be valid for any Reggeon.\footnote{Other analytic properties of 
parametrizations (\ref{pomeron}) in no way deserve serious consideration. These expressions should be treated just as some nonanalytic quantitative approximations (valid 
at low negative $t$ only) to the corresponding true dynamic functions whose analytic structure is still unknown.}
\begin{table}[ht]
\begin{center}
\begin{tabular}{|l|l|}
\hline
\bf Parameter          & \bf Value                   \\
\hline
$\alpha_{\rm P}(0)-1$  & $0.109\pm 0.017$            \\
$\tau_a$               & $(0.535\pm 0.057)$ GeV$^2$  \\
$g_{pp\rm P}(0)$       & $(13.8\pm 2.3)$ GeV         \\
$a_g$                  & $(0.23\pm 0.07)$ GeV$^{-2}$ \\
\hline
\end{tabular}
\end{center}
\vskip -0.2cm
\caption{The parameter values for (\ref{pomeron}) obtained via fitting to the elastic scattering data.}
\label{tab1}
\end{table}

Concerning the 3P-coupling, in this eprint we restrict ourselves by the simplest approximation 
$g_{3\rm P}(t_1,t_2,t_3)\approx g_{3\rm P}(0,0,0)\equiv g_{3\rm P}$. Such a rough variant of the 3P-model for the high-energy SD of nucleons is strongly correlated with the 
Pomeron exchange approximation for the nucleon-nucleon elastic scattering, since $g_{3\rm P}$ determines only the absolute value of the SD observables.\footnote{A case of 
quite different dynamical structure of the Regge residue is discussed in \cite{ryutin}.} Our ignoring of the nontrivial analytic structure of $g_{3\rm P}$ is justified by 
the model estimation of the exponential $t$-slope $B$ of the SD differential cross-section at $\sqrt{s}=1.8$ TeV and\linebreak 0.05 GeV$^2\le -t\le 0.11$ GeV$^2$, 
$B^{model}=11.7$ GeV$^{-2}$, which is quite consistent with the measured value \cite{E7102}: $B^{E710}=(10.5\pm 1.8)$ GeV$^{-2}$.

\begin{figure}[ht]
\vskip -0.3cm
\begin{center}
\epsfxsize=8.2cm\epsfysize=8.2cm\epsffile{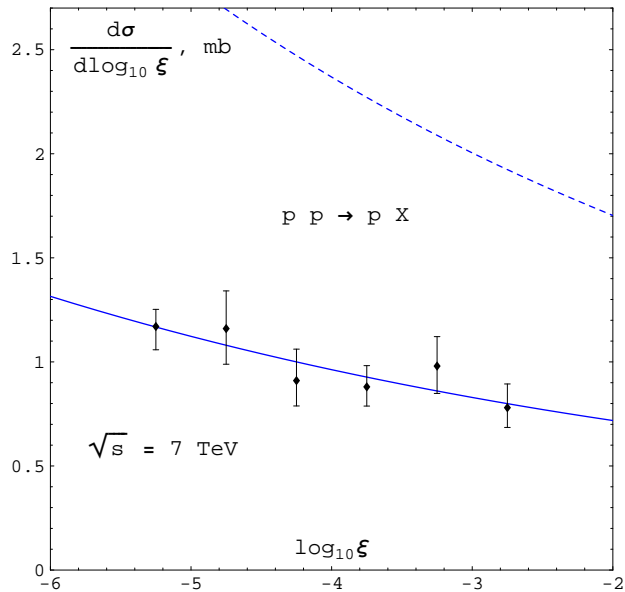}
\end{center}
\vskip -0.7cm
\caption{The 3P-model predictions (the solid line) for $g_{\rm 3P}=0.64$ GeV {\it versus} the CMS data \cite{CMS}. The dashed line corresponds to the cross-section without 
the $pp$ absorptive corrections.}
\label{cms}
\end{figure}

To obtain the value of $g_{3\rm P}$ we need just to normalize our variant of the 3P-model to the experimental data in that kinematic range wherein the total contribution of 
secondary Reggeon interactions is expected to be negligible. The recent CMS data \cite{CMS} make it possible\linebreak (see Fig. \ref{cms}): $g_{\rm 3P}=0.64$ GeV. The 
relative uncertainty of this quantity is not higher than the CMS data normalization uncertainty (10\% or even less). 

The comparative weakness of the 3P-interaction ($g_{3\rm P}/g_{pp\rm P}(0)\sim 0.05$) allows to neglect the absorptive corrections for the effective Pomeron-proton scattering 
within the relevant interval of $M_X$. Namely, the contribution of these corrections into the SD observables is much less than the total experimental uncertainties of the SD 
cross-sections. Contrary to ${\rm P}p$ scattering, the account of the nucleon-nucleon absorption (taken by introduction of factors $e^{i\delta(s,b_1)}$ and 
$e^{-i\delta^*(s,b_2)}$ in (\ref{diffcross})) is crucial since it significantly lowers the cross-section value and allows to reproduce successfully the $\xi$-behavior of 
$\frac{d\sigma}{d\log_{10}\xi}$ (compare the solid and dashed lines in Fig. \ref{cms}).\\ Certainly, the used approximation is invalid at asymptotically high values of $M_X$, 
since in the asymptotic regime {\it all} the absorptive corrections must be taken into account to satisfy the unitarity condition. A discussion of the unitarity 
problem can be found in \cite{fiore} and \cite{martynov}.

\begin{figure}[ht]
\vskip -0.2cm
\epsfxsize=8.2cm\epsfysize=8.2cm\epsffile{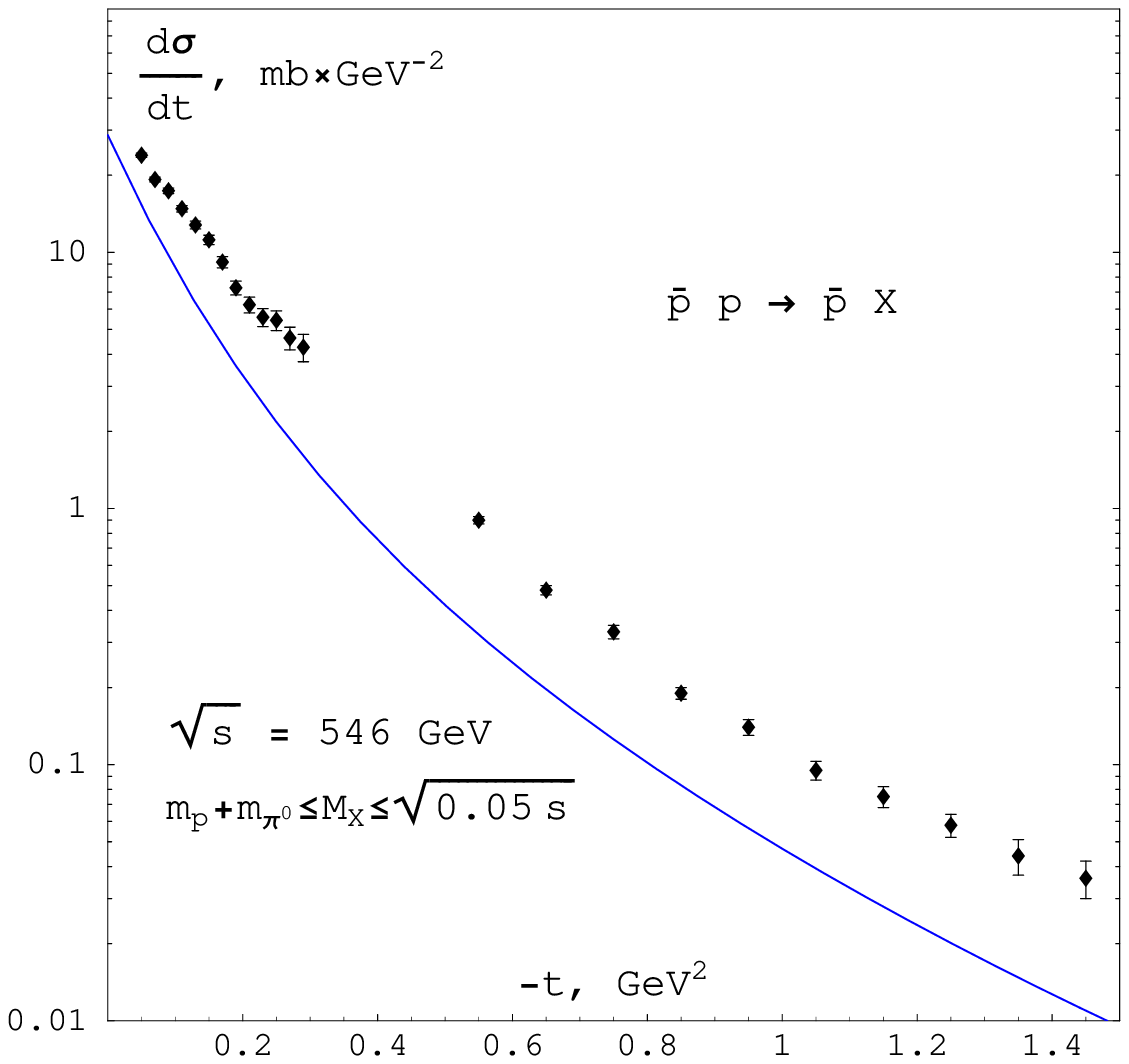}
\vskip -8.3cm
\hskip 8.75cm
\epsfxsize=8.25cm\epsfysize=8.25cm\epsffile{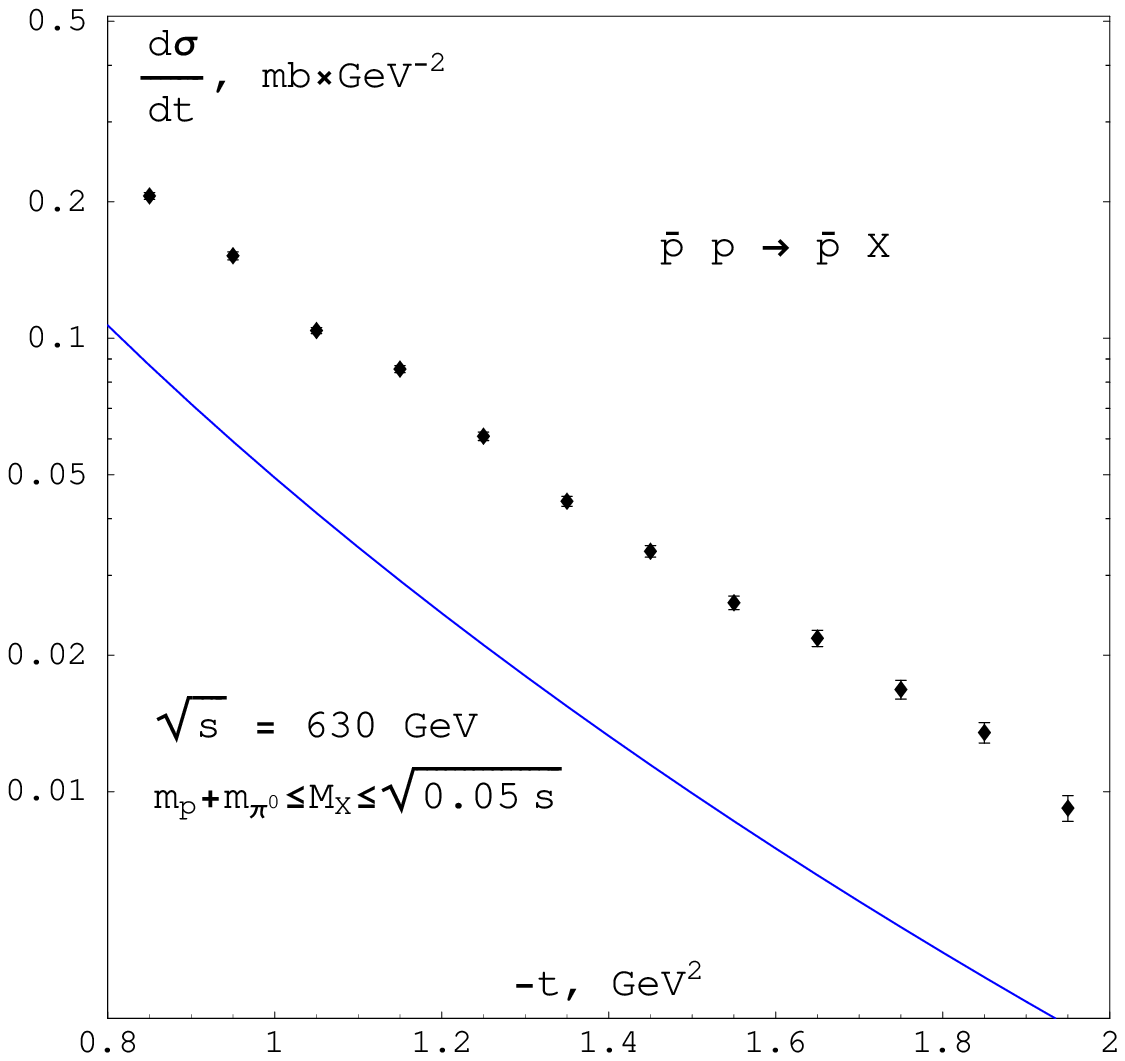}
\epsfxsize=8.3cm\epsfysize=8.3cm\epsffile{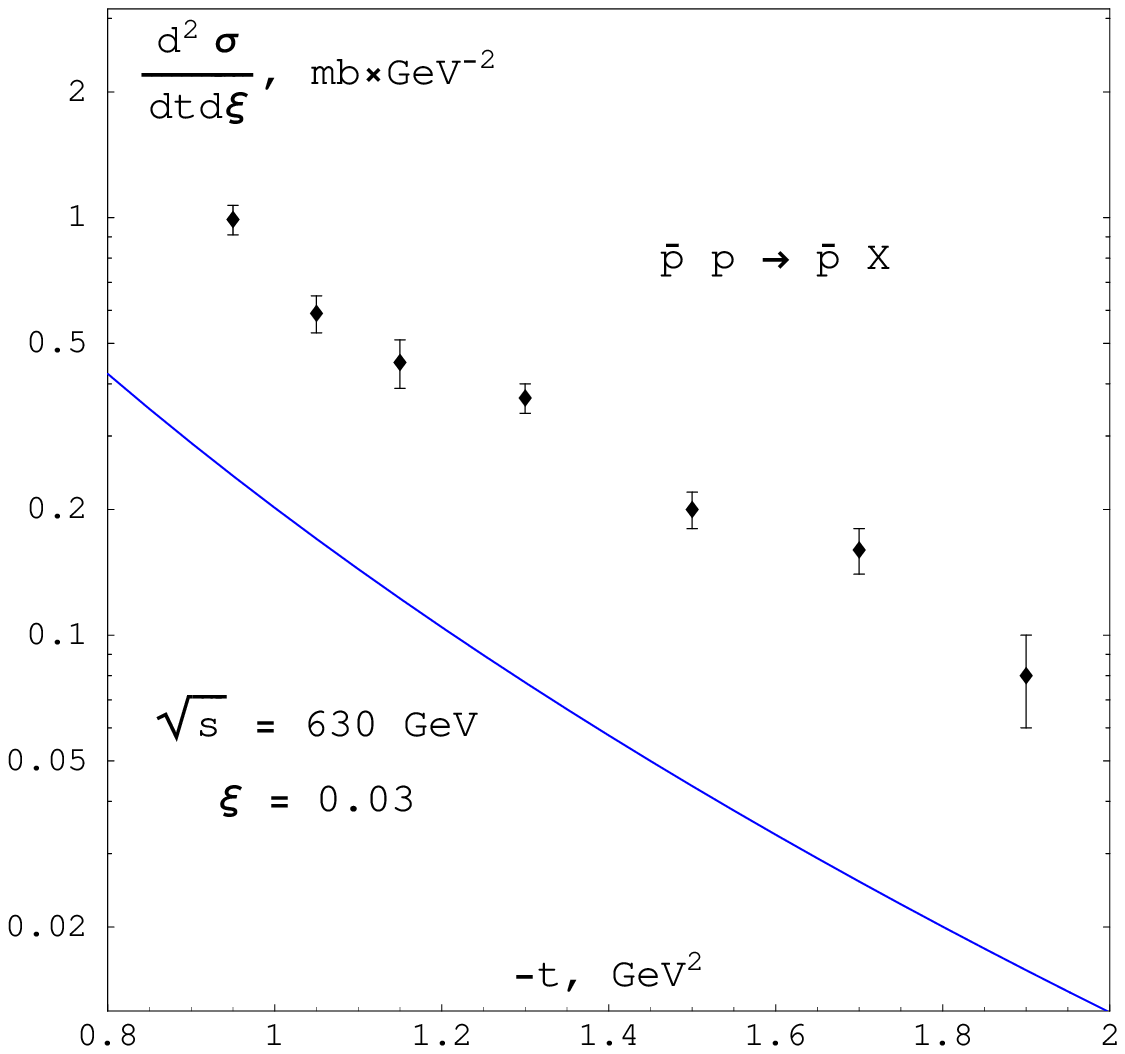}
\vskip -8.4cm
\hskip 8.7cm
\epsfxsize=8.55cm\epsfysize=8.55cm\epsffile{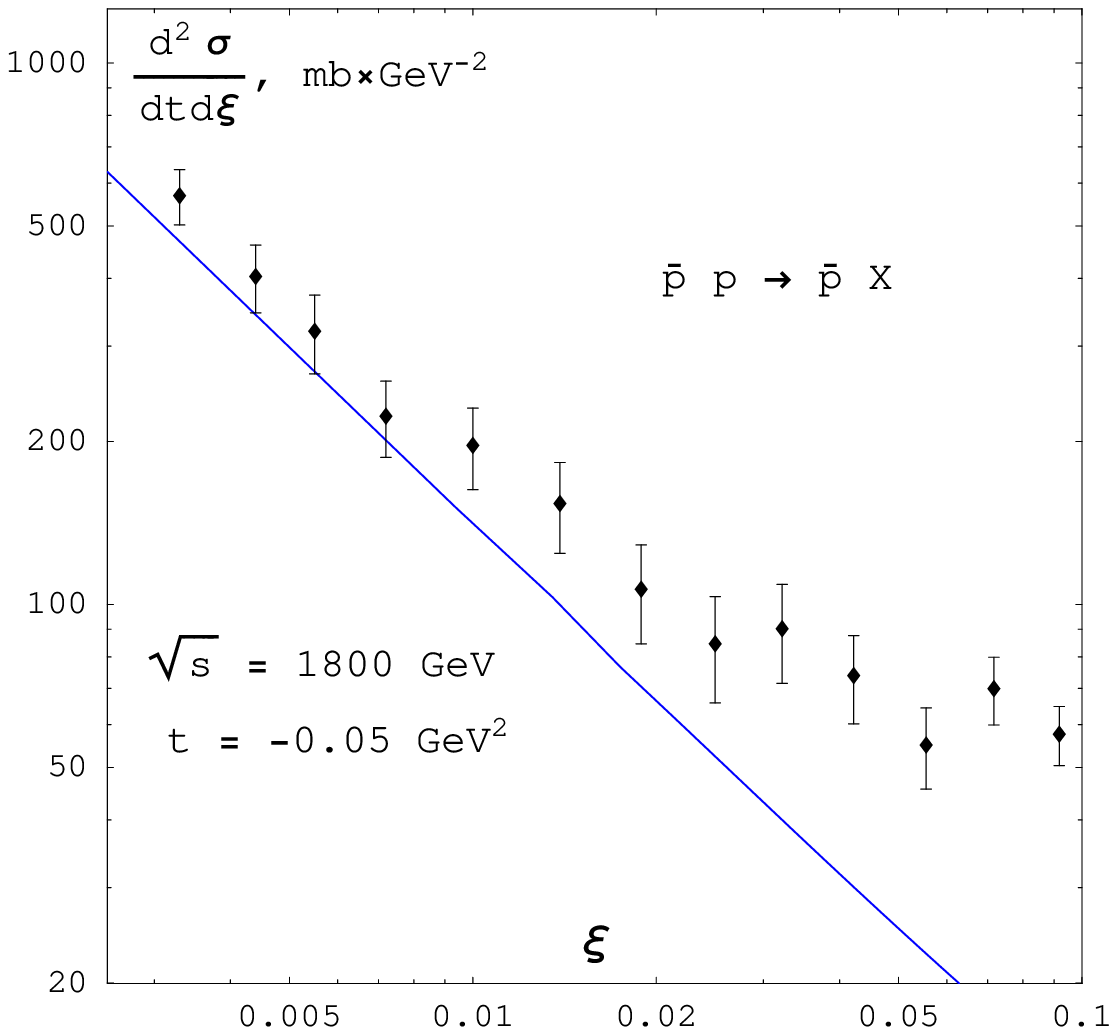}
\vskip -0.4cm
\caption{The 3P-model predictions for $g_{\rm 3P}=0.64$ GeV {\it versus} the UA4 \cite{UA4}, UA8 \cite{UA8}, and CDF \cite{montanha} data on the SD differential 
cross-sections.}
\label{diff}
\end{figure}

Comparison of the 3P-model outcomes with other available data on the differential distributions \cite{UA4,UA8,montanha} and the integrated cross-section 
$\sigma_{\rm SD}=2\int\int\frac{d^2\sigma}{dt dM_X^2}dtdM_X^2$ \cite{E7102,UA4,UA5,E7101,CDF,ALICE} (see Fig. \ref{diff} and Table \ref{integr}) reveals an absolute 
incompatibility of the considered approximation with the SD observables at the SPS energies. This divergence is the main subject of further discussion.

\newpage

\section*{Discussion}

\begin{figure}[ht]
\vskip -0.9cm
\epsfxsize=8.2cm\epsfysize=8.2cm\epsffile{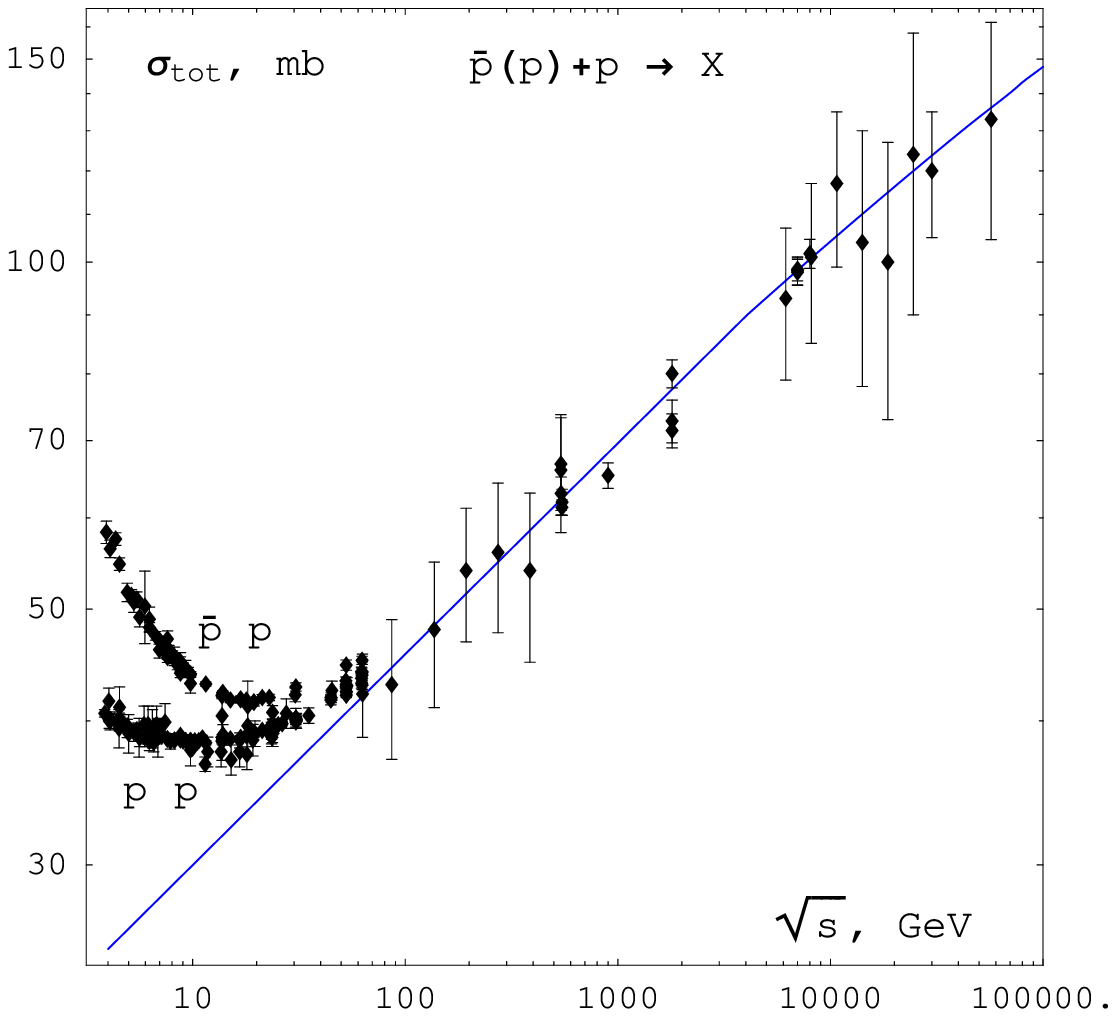}
\vskip -7.85cm
\hskip 8.9cm
\epsfxsize=7.45cm\epsfysize=7.45cm\epsffile{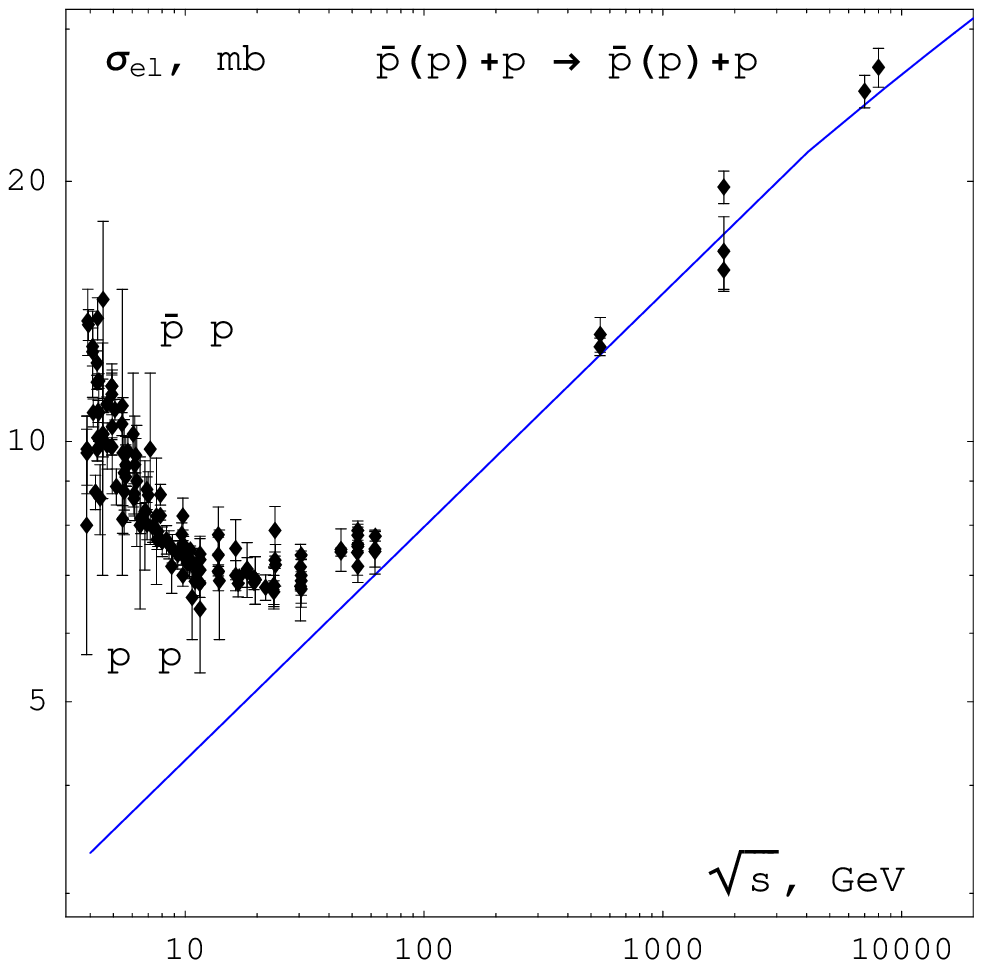}
\vskip -0.1cm
\caption{The Pomeron-exchange approximation \cite{godizov} {\it versus} the data on the nucleon-nucleon total (left) and elastic (right) cross-sections \cite{cross}.}
\label{tot}
\end{figure}

To understand if the 3P-model underestimation of the SD cross-sections at the SPS energies could be interpreted in terms of secondary Reggeon contributions let us consider 
the collision energy dependence of the nucleon-nucleon total and elastic cross-sections (Fig. \ref{tot}). Regarding these observables, the whole interval of available 
$\sqrt{s}$ can be divided into three parts: the region of the Pomeron absolute dominance ($\sqrt{s}>$ 100 GeV), the region of the Pomeron relative dominance 
($\sqrt{s}>$ 14 GeV), and the region of the Pomeron exchange approximation irrelevance ($\sqrt{s}<$ 14 GeV). Concerning SD itself, the dynamic structure 
(\ref{triple})---(\ref{eik}) of $\frac{d^2\sigma}{dt dM_X^2}$ points to the fact that the 3P-model validity is conditioned rather by low values of $\xi$ and high values of 
$M_X$ than by high value of $\sqrt{s}$ directly. Therefore, by qualitative analogy with the proton-(anti)proton elastic scattering, it is justified to single out 
conventionally the following kinematic ranges of SD: the region of the 3P-interaction absolute dominance, $\frac{10^4\,{\rm GeV^2}}{s}<\xi<2\cdot 10^{-4}$, and the region of 
the 3P-interaction relative dominance, $\frac{200\,{\rm GeV^2}}{s}<\xi<0.01$. The former range is null at the SPS and Tevatron. The latter one is null at the ISR and lower 
energies.

As the observed values of $\sigma^{(\bar p p)}_{tot}$ and $\sigma^{(\bar p p)}_{el}$ at $\sqrt{s}=$ 5 GeV are, respectively, twice and 3-4 times higher than the Pomeron 
exchange model \cite{godizov} predictions, so it seems quite natural that the total contribution of secondaries into the $\bar p p$ SD differential cross-section is about 
75-80\% at $\sqrt{s}=$ 630 GeV and $\xi=$ 0.03 (see Fig. \ref{diff}). 
\begin{table}[ht]
\begin{center}
\begin{tabular}{|l|l|l|l|}
\hline
\bf Experiment    & \bf Kinematic range  &  $\sigma^{\rm exp}_{\rm SD}(s)$, mb & $\sigma^{\rm model}_{\rm SD}(s)$, mb \\
\hline
UA4   \cite{UA4}  & $\sqrt{s}=546$  GeV,    $m_p+m_{\pi^0}<M_X<\sqrt{0.05\,s}$ &   9.4  $\pm$ 0.7           &  5.3    \\
UA4   \cite{UA4}  & $\sqrt{s}=546$  GeV,            4 GeV$<M_X<\sqrt{0.05\,s}$ &   6.4  $\pm$ 0.4           &  3.5    \\
CDF   \cite{CDF}  & $\sqrt{s}=546$  GeV  $\sqrt{1.4}$ GeV$<M_X<\sqrt{0.15\,s}$ &   7.89 $\pm$ 0.33          &  5.6    \\
CDF   \cite{CDF}  & $\sqrt{s}=546$  GeV,            4 GeV$<M_X<\sqrt{0.15\,s}$ &   5.4  $\pm$ 0.3           &  3.9    \\
UA5   \cite{UA5}  & $\sqrt{s}=0.9$  TeV,    $m_p+m_{\pi^0}<M_X<\sqrt{0.05\,s}$ &   7.8  $\pm$ 1.2           &  6.5    \\
ALICE \cite{ALICE}& $\sqrt{s}=0.9$  TeV,    $m_p+m_{\pi^0}<M_X<200$ GeV        & $11.2^{+1.6}_{-2.1}$       &  6.5    \\
E-710 \cite{E7101}& $\sqrt{s}=1.8$  TeV,    $m_p+m_{\pi^0}<M_X<\sqrt{0.05\,s}$ &  11.7  $\pm$ 2.3           &  8.5    \\
E-710 \cite{E7102}& $\sqrt{s}=1.8$  TeV    $\sqrt{2}$ GeV$<M_X<\sqrt{0.05\,s}$ &   8.1  $\pm$ 1.7           &  8.0    \\
CDF   \cite{CDF}  & $\sqrt{s}=1.8$  TeV, $\sqrt{1.4}$ GeV$<M_X<\sqrt{0.15\,s}$ &   9.46 $\pm$ 0.44          &  8.9    \\
ALICE \cite{ALICE}& $\sqrt{s}=2.76$ TeV,    $m_p+m_{\pi^0}<M_X<200$ GeV        & $12.2^{+3.9}_{-5.3}$       &  8.8    \\
ALICE \cite{ALICE}& $\sqrt{s}=7$    TeV,    $m_p+m_{\pi^0}<M_X<200$ GeV        & $14.9^{+3.4}_{-5.9}$       & 11.1    \\
\hline
\end{tabular}
\caption{The 3P-model predictions for $g_{\rm 3P}=0.64$ GeV {\it versus} available data on the SD integrated cross-section 
$\sigma_{\rm SD}=2\int\int\frac{d^2\sigma}{dt dM_X^2}dtdM_X^2$.}
\label{integr}
\end{center}
\end{table}

As well, at low values of $M_X$, the impact of $f$-Reggeon exchanges on the effective Pomeron-proton scattering might be significant (the PP$f$-vertex), and, certainly, 
the diffractive excitation of proton to various resonance states noticeably raises the SD cross-section value in the range $M_X<$ 4 GeV (for details, see
\cite{jenkovszky}). However, these effects are important in the limited interval of $M_X$. Consequently, at higher energies, their relative contribution into 
$\sigma_{\rm SD}$ decreases, as the upper bound of the $M_X$-range of the 3P-interaction dominance grows fast with the collision energy increasing. This is the reason 
why the 3P-model relative underestimation of the cross-sections integrated over the whole kinematic range of SD at $\sqrt{s}=$ 1800 GeV is much smaller than at 
$\sqrt{s}=$ 546 GeV (see Table \ref{integr}).

In view of the above-presented qualitative analysis, we come to the main conclusions:
\begin{itemize}
\item The 3P-model (\ref{triple})---(\ref{eik}) with constant 3P-vertex yields satisfactory predictions for such SD observables as the $t$-slope of $d\sigma/dt$ at 
$\sqrt{s}=$ 1.8 TeV and 0.05 GeV$^2<-t<0.11$ GeV$^2$ \cite{E7102} and the $\xi$-behavior of $d\sigma/d\xi$ at $\sqrt{s}=$ 7 TeV and $10^{-5.5}<\xi<10^{-2.5}$ \cite{CMS}. So 
and thus, this simple approximation allows to obtain a well-grounded estimation of the triple-Pomeron coupling value and is expected to be quite applicable ({\it i.e.}, to be 
a reliable phenomenological tool) within the above-mentioned $\xi$-intervals of the 3P-interaction absolute and relative dominance. 
\item The 3P-model underestimation of the measured SD cross-sections at the SPS and Tevatron energies gains a natural explanation in terms of secondary Reggeon exchanges 
and resonance contributions.
\end{itemize}

\section*{Acknowledgments} The author is grateful to V.A. Petrov and R.A. Ryutin for numerous discussions.

\end{document}